\documentclass[a4paper,12pt]{article}
\usepackage{geometry}
\usepackage[cp1251]{inputenc}
\usepackage{epsfig}
\usepackage{graphicx}

\begin{document}

\section*{
\begin{center}
Wind Dynamics and Circumstellar Extinction Variations in the T~Tauri Star RY~Tau
\end{center}
}

\begin{center}
{\large Elena V. Babina, Svetlana A. Artemenko and Peter P. Petrov}
\end{center}

\begin{center}
\it{Crimean Astrophysical Observatory}
\end{center}

\subsection*{Abstract}
The wind interaction with the dusty environment of the classical T Tauri star RY~Tau has been investigated. During two seasons of 2013-2015  we carried out a spectroscopic and photometric ($BVR$) monitoring of the star. A correlation between the stellar brightness and the radial velocity of the wind determined from the H$\alpha$ and Na D line profiles has been found for the first time.
The irregular stellar brightness variations are shown to be caused by extinction in a dusty disk wind at a distance of about 0.2 AU from the star. We suppose, that  variations of the circumstellar extinction results from  cyclic rearrangements of the stellar magnetosphere and coronal mass ejections, which affect the dusty disk wind near the inner boundary of the circumstellar disk.
%% Similarity and differences between RY Tau,  AA Tau and UXors are discussed.

\subsection*{Introduction}

Young low-mass ($\leq$ 2 M$_\odot$) stars with accretion disks are called classical T Tauri stars. The age of these stars is $\sim$1 Myr. At this age, the planetary system is in the process of formation, the star is surrounded by a disk of gas and dust, and the accretion of disk material
onto the star continues. T Tauri stars have a deep convective envelope and quite a strong surface magnetic field ($\sim$1 kG). The disk accretion of conducting material is stopped by
the magnetic field at a distance of several stellar radii, and the gas streams are channeled along magnetic field lines onto the stellar surface. This is the socalled magnetospheric accretion (Camenzind 1990; Koenigl 1991; Bouvier et al. 2007). The observational evidence for magnetospheric accretion in T Tauri stars is well known: the Doppler shift of absorption
lines formed in the infalling gas, and the hot spots on the stellar surface at the base of the accretion columns. The accretion rate ranges from 10$^{-10}$ to 10$^{-7}$ M$_\odot$ yr$^{-1}$ (Hartigan et al. 1995; Gullbring et al. 1998; White and Ghez 2001).

The radius of the boundary region between the accretion disk and 
stellar magnetosphereis is approximately equal to
the corotation radius (the distance from the star at
which the angular velocity of the star is equal to the
Keplerian angular velocity in the disk). For a solar-mass
star rotating with a period of 7 days (the mean
rotation period of T Tauri stars), the corotation radius
is equal to five stellar radii. Although the accretion
disk is composed of gas and dust, dust cannot be too
close to the star. The inner radius of the dust disk
is determined by the dust sublimation temperature.
Most of the T Tauri stars are of spectral type K, and
the inner radius of the dust disk is  $\sim$0.1 AU. Since
hot dust radiates in the near infrared, the inner radii of
the dust disks are measured by infrared interferometry
(see, e.g., Akeson et al. 2005).

Apart from the accretion of matter, classical
T Tauri stars exhibit an intense wind. Various types
of wind are possible. The stellar wind (an analog
of the solar one) flows along open magnetic field
lines predominantly from polar regions (Matt and
Pudrits 2005). The disk wind flows from the disk surface.
The disk has its own magnetic field; therefore,
the magneto-centrifugal effect, the acceleration of
ionized gas along open magnetic field lines in the disk,
arises (Blandford and Payne 1982). Other types of
wind can arise at the magnetosphere–disk boundary:
an X-wind (Shu et al. 1994) and a conical wind
(Romanova et al. 2009). Below, we will arbitrarily
call this a “magnetospheric” wind. Episodic coronal
mass ejections are also possible (Ferreira et al. 2000).
The difference between the profiles of emission
lines forming in the stellar and disk winds as well as
their dependence on the inclination of the stellar rotation
axis were discussed by Kurosawa et al. (2011)
and Grinin and Tambovtseva (2011). The profiles
of emission lines forming in the conical wind were
considered by Kurosawa and Romanova (2012).

Significantly, the accretion and wind processes are
nonstationary, which affects the emission
line profiles. Periodic variations due to rotational
modulation in the case of an axisymmetric
magnetic field are also occasionally observed (Petrov
et al. 2001; Bouvier et al. 2003; Petrov et al. 2011).
The most dramatic phenomena occur at the base
of themagnetospheric wind. The difference in angular
velocity between the star and the disk at its inner
boundary causes the magnetic loops connecting the
star to the disk to be twisted, leading to an amplification
of the magnetic energy, the opening (inflation)
of the magnetosphere, and the ejection of material in
the form of a wind, whereupon the magnetosphere is
again restored. Such cyclic rearrangements of the
magnetosphere and mass ejections can occur quasiperiodically
on a time scale of several stellar rotations
(Goodson et al. 1997; Romanova et al. 2009).

In this study, we are interested in the interaction of
a nonstationary magnetospheric wind with the dusty
environment of the star. Large dust grains are concentrated
in the disk plane, but fine dust is present in
the disk atmosphere and can be carried away by the
disk wind through the collisions of dust grains with
neutral gas atoms (Safier 1993). The calculations of
the thermal balance of dust performed by Tambovtseva
and Grinin (2008) showed the dust to survive in
the hot wind from young stars. If the line of sight to
the star passes at a small angle to the disk plane, then
the dust attenuates appreciably the light from the star
(circumstellar extinction).

The disk wind lifting dust above the disk plane
is more intense at the inner, hottest, disk boundary.
This creates a kind of a dust screen around the star.
Magnetospheric wind variations can exert some influence
on this screen. In T Tauri stars, the distance
from the magnetosphere-disk boundary (near
the corotation radius) to the inner boundary of the
dust disk (dust sublimation radius) is relatively small:
$\approx$\,0.1 AU. During mass ejections from the magnetosphere
with a velocity of 200 -- 300 km\,s$^{-1}$, this
distance is traversed in about one day. Consequently,
one might expect a correlation between the observed
wind velocity and circumstellar extinction, possibly,
with some time delay.

To investigate the wind dynamics and the possible
influence of the wind on circumstellar extinction, we
carried out a long series of spectroscopic and photometric
observations for the T Tauri star RY Tau. We
chose this object for our observations, because the
circumstellar disk of RY Tau is inclined at an angle of
20\,$^\circ \pm 5\,^\circ$ degrees to the line of sight (Isella et al. 2010),
i.e., we see the star through the disk wind.

The main parameters of RY Tau (see Lopez-Martinez and Gomez de Castro 2014) are: spectral type G1, bolometric luminosity $L$ = 9.6 L$\odot$, radius
$R$ = 2.9 R$\odot$, and mass $M$ = 2 M$\odot$. On the log $T$ --
log $L$ diagram (Siess et al. 2000), the star is on a
radiative track at the boundary between the T Tauri
and Herbig Ae stars. The broad photospheric lines,
$v\,\sin i$ $\approx$\,50 km\,s$^{-1}$, point to a rapid rotation of RY Tau.
At this stellar radius and a high inclination of the
rotation axis, the rotation period of the star can be within
the range 2.3 -- 3.4 days.

The spectral energy distribution for the star in the
range from 0.1 to 1000 $\mu$m corresponds to the model
of a disk of gas and dust with a mass of $\approx$\,0.03 M$\odot$
(Robitaille et al. 2007). According to infrared interferometry
(Akeson et al. 2005; Pott et al. 2010;
Schegerer et al. 2008), the inner radius of the dust
disk is $R_{in}$ = 0.2 -- 0.3 AU, depending on the disk
model. A millimeter-band disk image shows that the
inclination of the disk rotation axis to the line of sight
is 65\,$^\circ$ -- 75\,$^\circ$ (Isella et al. 2010). RY Tau can be
classified by this parameter as belonging to UX Ori
stars (Grinin et al. 1991).

RY Tau has an extended bipolar jet (St-Onge and
Bastien 2008). The jet structure and kinematics
were investigated by Coffey et al. (2015) and Agra-Amboage et al. (2009). 
X-ray emission from the jet of RY Tau was observed by Chandra (Skinner
et al. 2011). In the optical range, RY Tau has an
emission line spectrum typical of T Tauri stars. The
variability of emission line profiles was investigated by
Chou et al. (2013).
RY Tau is one of the brightest T Tauri stars, and its
photometric history has been well documented. The
most detailed analysis of the photometry for RY Tau
was performed by Zaitseva (2010) based on the data
obtained from 1965 to 2000. Quasi-periodic variations
that are probably associated with eclipses by
dust clouds in the circumstellar disk were revealed.
Periods of 7.5, 20, and 20.9 days manifested themselves
in different years of observations. No stable
period that could be identified with the rotation period
of the star was detected. Continuous photometry for
RY Tau during three weeks by the MOST satellite did
not reveal a periodic signal either (Siwak et al. 2011).

Two noticeable brightenings, in 1983–1984 and
1996–1997, were observed against the background
of irregular stellar brightness fluctuations in the range
$V=9^m$.5 -- $11^m$.5 (Herbst and Stine 1984; Herbst
et al. 1994; Zaitseva et al. 1996). The color–magnitude diagram 
for RY Tau is typical of UX Ori stars. As the star fades, its $B - V$ color index
initially increases and then, at a deep minimum,
decreases, though a large scatter of individual color
indices is observed. The spectral characteristics of
the star do not change with brightness. This suggests
that circumstellar extinction variations are mainly
responsible for the variability (Petrov et al. 1999).

\subsection*{Observations}

The observations were carried out at the Crimean
Astrophysical Observatory during two seasons
from 2013 to 2015. The spectroscopic observations
were carried out with the echelle spectrograph of
the 2.5-m Shajn telescope using an iKON-L 936
(2048 x 2048 pixel) CCD camera. With a 2\,$^{\prime\prime}$ entrance
slit, the spectral resolution was $\lambda/\Delta\lambda$ $\approx$ 25 000. 
The spectroscopic observations were performed in short
series  of four days to cover the rotation period of the star 
(about three days) and to reveal the possible rotational modulation 
of emission lines.
Because of the weather conditions, these series were
sometimes shorter. As a rule, several 30-min spectral
exposures were taken over the night, which were then
summed. The mid-exposure times of observations
are given in Table~1. In total, we observed the star
for 32 nights: 12 and 20 nights in the 2013--2014 and
2014--2015 seasons, respectively.
                                                           
The photometric observations of RY Tau were carried
out at the 1.25-m AZT-11 telescope using two
detectors: (1) an FLI ProLine PL230 (2048 x 2048
pixel) CCD camera with a 10.\,$^\prime$9 x 10.\,$^\prime$9 field of view
and an angular resolution of 0.$^{\prime\prime}$32/pixel and (2) a
Finnish five-channel pulse-counting photometer–polarimeter. 
The observations were performed in
Johnson’s $BVR$ bands. The root-mean-square error
of each measurement was, on average, 0$^m$.005 and 0$^m$.01 for the photoelectric and CCD observations,
respectively. The results are presented in Table~1. In
most cases, the times of our photometric observations
coincided with the mean times of spectral exposures
within about one hour. Given that the brightness of
RY Tau changes, on average, by 0$^m$.1--0$^m$.2 in one
day, the error of the magnitudes listed in Table~1 
at the time of our spectroscopic observations can reach 0$^m$.01. 
When we failed to perform photometry simultaneously with spectroscopy, 
we determined the stellar brightness at the time of our spectroscopic
observations by interpolation between adjacent dates,
and the errors could reach 0$^m$.1. These errors are given in Table~1.

\begin{table}[h!]
\centering
\caption{Times of observations, stellar brightness, equivalent width of H$\alpha$ emission
 and radial velocity of outflow.}
\label{tab:tabl1}
\bigskip
\tabcolsep=3.0mm
\vspace{5mm}\begin{tabular}{|c|cc|cc|cc|} 
\hline
JD          & $V$   & $\sigma$  &  EW  & $\sigma$ & RV$_{max}$ & $\sigma$   \\
2450000+    & m     & m         & \AA  &  \AA     &   km/s     &   km/s    \\
\hline                           
   6592.500 & 10.46 &  0.01 &  18.3 & 1.0 & -146 & 10 \\
   6593.437 & 10.70 &  0.01 &  19.3 & 1.0 & -137 &  8 \\
   6594.371 & 10.45 &  0.01 &  11.5 & 0.5 & -130 &  7 \\
   6595.347 & 10.38 &  0.01 &   9.7 & 0.5 & -130 &  6 \\
   6605.535 & 10.09 &  0.01 &   7.1 & 1.5 & -185 &  6 \\
   6606.442 & 10.11 &  0.01 &   7.2 & 1.5 & -225 & 12 \\  
   6621.312 & 10.19 &  0.01 &   6.5 & 0.5 & -240 & 10 \\ 
   6691.387 &  9.80 &  0.05 &   8.1 & 0.5 & -232 &  6 \\ 
   6742.226 & 10.65 &  0.10 &  11.2 & 1.5 & -155 & 10 \\
   6743.187 & 10.70 &  0.10 &  12.9 & 2.0 & -175 &  5 \\
   6744.244 & 10.73 &  0.10 &  17.3 & 0.2 & -163 & 10 \\ 
   6748.294 & 10.97 &  0.10 &  21.4 & 1.5 & -160 & 10 \\ 
   6905.524 & 10.30 &  0.01 &   7.1 & 1.0 & -186 & 15 \\ 
   6906.520 & 10.26 &  0.01 &  11.0 & 1.0 & -220 & 10 \\ 
   6907.503 & 10.30 &  0.05 &   8.5 & 0.4 & -220 &  6 \\ 
   6908.508 & 10.35 &  0.01 &   9.4 & 0.3 & -196 &  8 \\ 
   6933.554 & 10.13 &  0.05 &  14.0 & 1.0 & -275 & 12 \\ 
   6934.485 & 10.10 &  0.05 &   8.8 & 0.3 & -275 & 12 \\ 
   6935.493 & 10.12 &  0.01 &  14.5 & 0.3 & -185 & 10 \\ 
   6936.499 & 10.17 &  0.01 &  11.2 & 0.3 & -196 &  8 \\ 
   6964.459 & 10.32 &  0.01 &   6.7 & 0.3 & -170 &  8 \\ 
   6975.437 & 10.42 &  0.01 &  16.2 & 0.5 & -209 &  6 \\ 
   6976.447 & 10.59 &  0.01 &  19.9 & 0.5 & -199 &  6 \\ 
   6977.462 & 10.60 &  0.05 &  15.0 & 0.5 & -200 & 10 \\ 
   6993.257 & 10.27 &  0.01 &  21.9 & 0.5 & -214 & 10 \\ 
   7058.219 & 10.62 &  0.01 &  22.0 & 0.5 & -173 & 15 \\ 
   7059.273 & 10.73 &  0.05 &   9.7 & 0.4 & -182 & 10 \\ 
   7060.238 & 10.80 &  0.10 &  12.2 & 0.5 & -176 & 15 \\ 
   7089.284 & 11.09 &  0.01 &  15.6 & 0.5 & -134 &  7 \\ 
   7090.245 & 11.18 &  0.01 &  17.6 & 0.5 & -134 &  7 \\ 
   7091.269 & 11.16 &  0.01 &  18.8 & 0.3 & -120 &  8 \\ 
   7092.267 & 11.15 &  0.01 &  15.7 & 0.5 & -125 &  8 \\ 
\hline                                       
\end{tabular}
\end{table}

\subsection*{Results} 

Over the time of our observations, the brightness
of RY Tau varied within the range from $V$ = 9$^m$.8 to $V$ = 11$^m$.2 without any noticeable dependence of the colors on brightness. Fig. 1 presents the light curve of RY Tau and the times of our spectroscopic observations.

%%%%%%%%%%%%%%%%%%%%%%%%%%%%%%%%%%%%%%%%%%%%%%%%%%%%%%%%
\begin{figure}[h]
\epsfxsize=14cm
\vspace{0.6cm}
\hspace{1cm}\epsffile{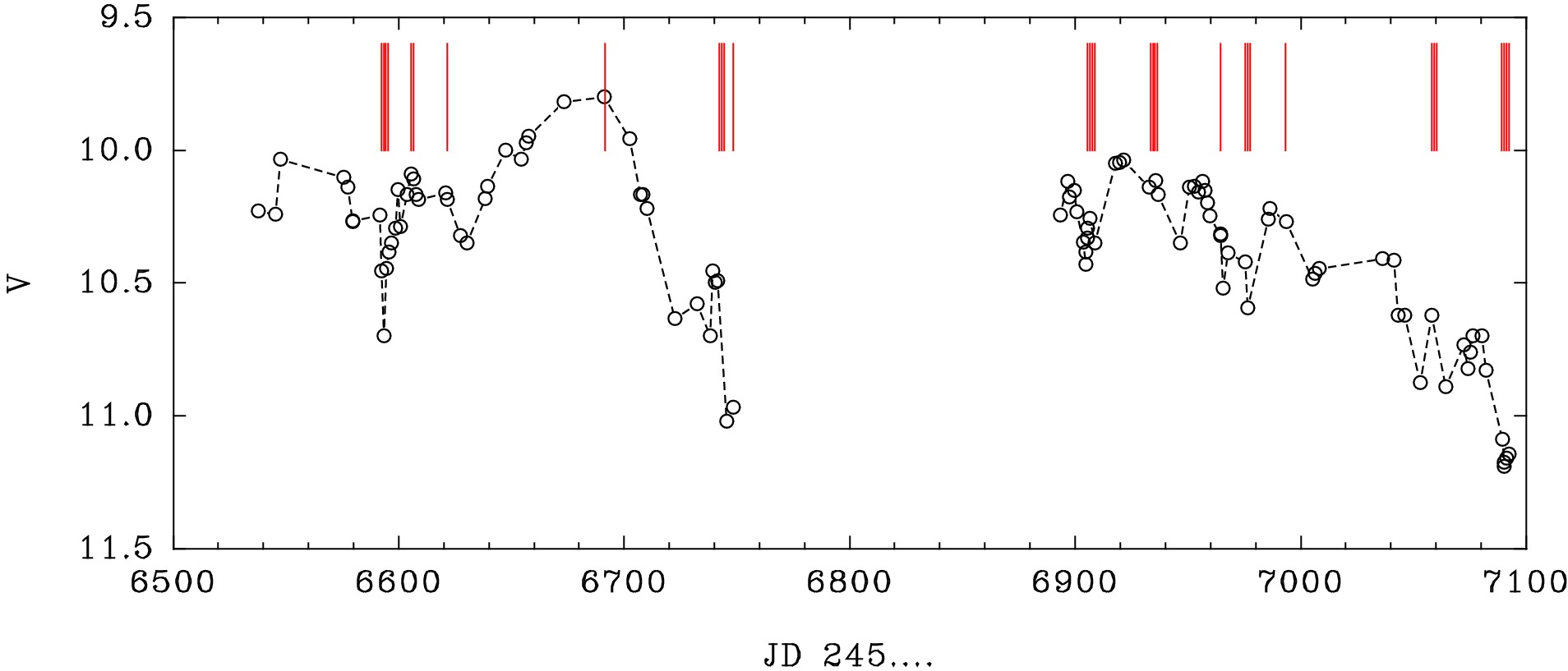}
\caption{\rm \footnotesize {Light curve of RY Tau in the 2013–-2014 and 2014–-2015 seasons. The vertical dashes mark the times of our spectroscopic observations. }}
\end{figure}
%%%%%%%%%%%%%%%%%%%%%%%%%%%%%%%%%%%%%%%%%%%%%%%%%%%%%%%%
%%%%%%%%%%%%%%%%%%%%%%%%%%%%%%%%%%%%%%%%%%%%%%%%%%%%%%%%
\begin{figure}[h]
\epsfxsize=6.0cm
\vspace{0.6cm}
\hspace{4cm}\epsffile{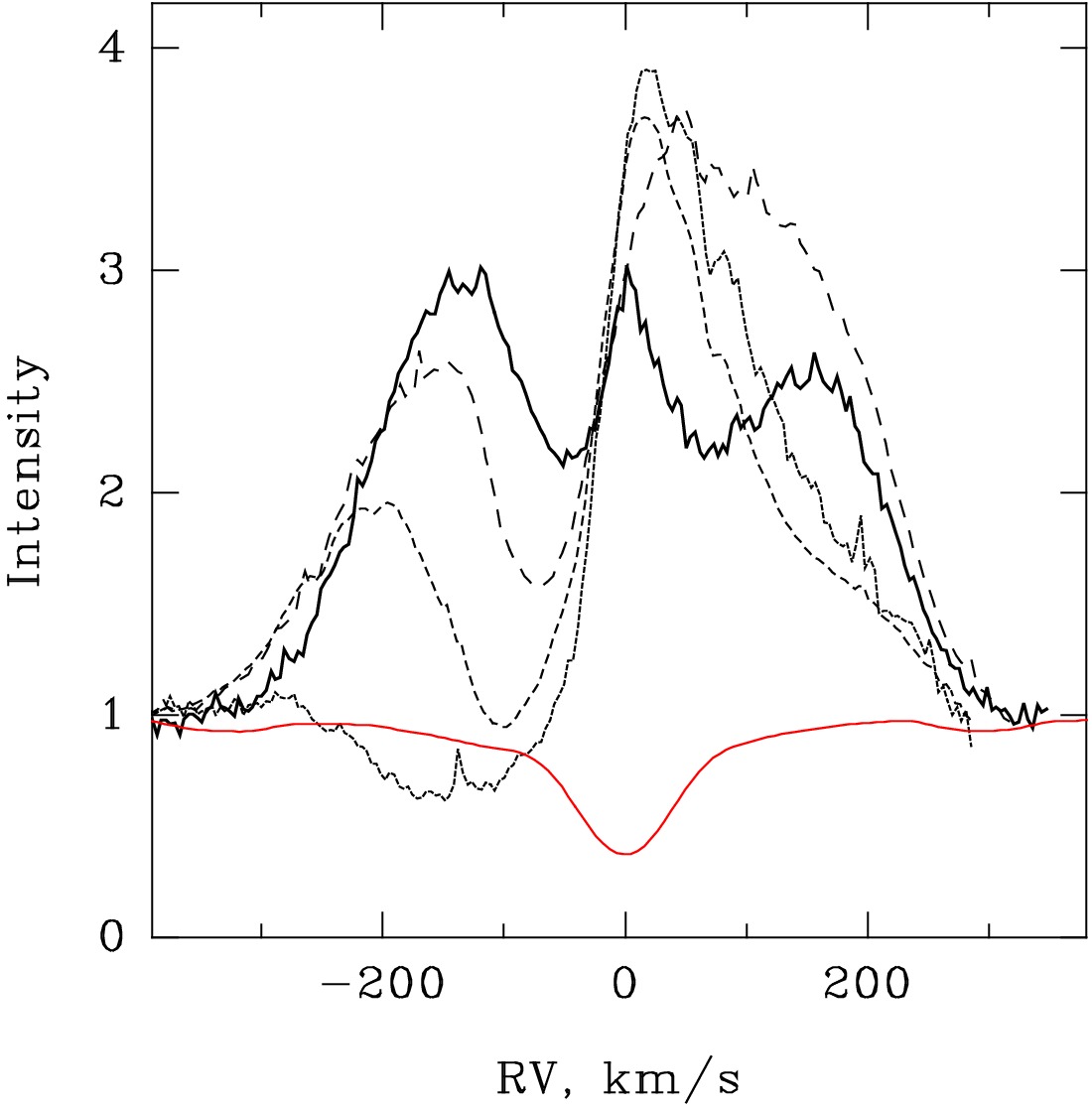}
\caption{\rm \footnotesize {Sample of H$\alpha$ profiles showing the range of radial velocity variability of the wind. The thin solid line at the bottom indicates
the photospheric profile of the star 35 Leo (G1.5 IV-V) rotationally broadened to 
$v\,\sin i$ = 50 km\,s$^{-1}$  }}
\end{figure}
%%%%%%%%%%%%%%%%%%%%%%%%%%%%%%%%%%%%%%%%%%%%%%%%%%%%%%%%%%

In this paper, we use the H$\alpha$ and Na D line profiles
as indicators of gas flows -- the wind and accretion.
The typical H$\alpha$ line profile consists of a
broad emission with symmetric wings extended to $\pm$300 km\,s$^{-1}$ 
and a blueshifted absorption indicating an outflow of matter. 
According to the classification of Reipurth et al. (1996), this is a type II B profile. 
This type of profile is characteristic of a disk wind when
the line of sight passes at a small angle to the disk
surface (Grinin and Tambovtseva 2011; Kurosawa
et al. 2011). If, however, the star is seen pole-on,
then the line of sight passes through the stellar wind
regions and a P Cyg (type IV B) profile is observed.
Fig. 2 presents a sample of several H$\alpha$ line profiles
in the spectrum of RY Tau. What is unusual is that
both types of profile with a gradual transition from
type II B to type IV B are observed. Obviously, this
is due to a change in wind geometry rather than a
change in inclination. A P Cyg profile suggests a
radial outflow with acceleration, which is probably
related to some temporary gas ejection episodes.

In addition, a dip in the red line wing, occasionally
fairly deep but not lower than the continuum, often
appears in the H$\alpha$ profile. Concurrently, an additional
absorption appears in the Na D profile on the red
side, indicating the infall of gas to the star with a
radial velocity up to +200 km\,s$^{-1}$. Signatures of both
outflow and accretion (type II B + II R) are often
observed simultaneously in the profiles of these lines.

%%%%%%%%%%%%%%%%%%%%%%%%%%%%%%%%%%%%%%%%%%%%%%%%%%%%%%%%%%%          
\begin{figure}[h!]
\epsfxsize=15cm
\vspace{0.6cm}
\hspace{1cm}\epsffile{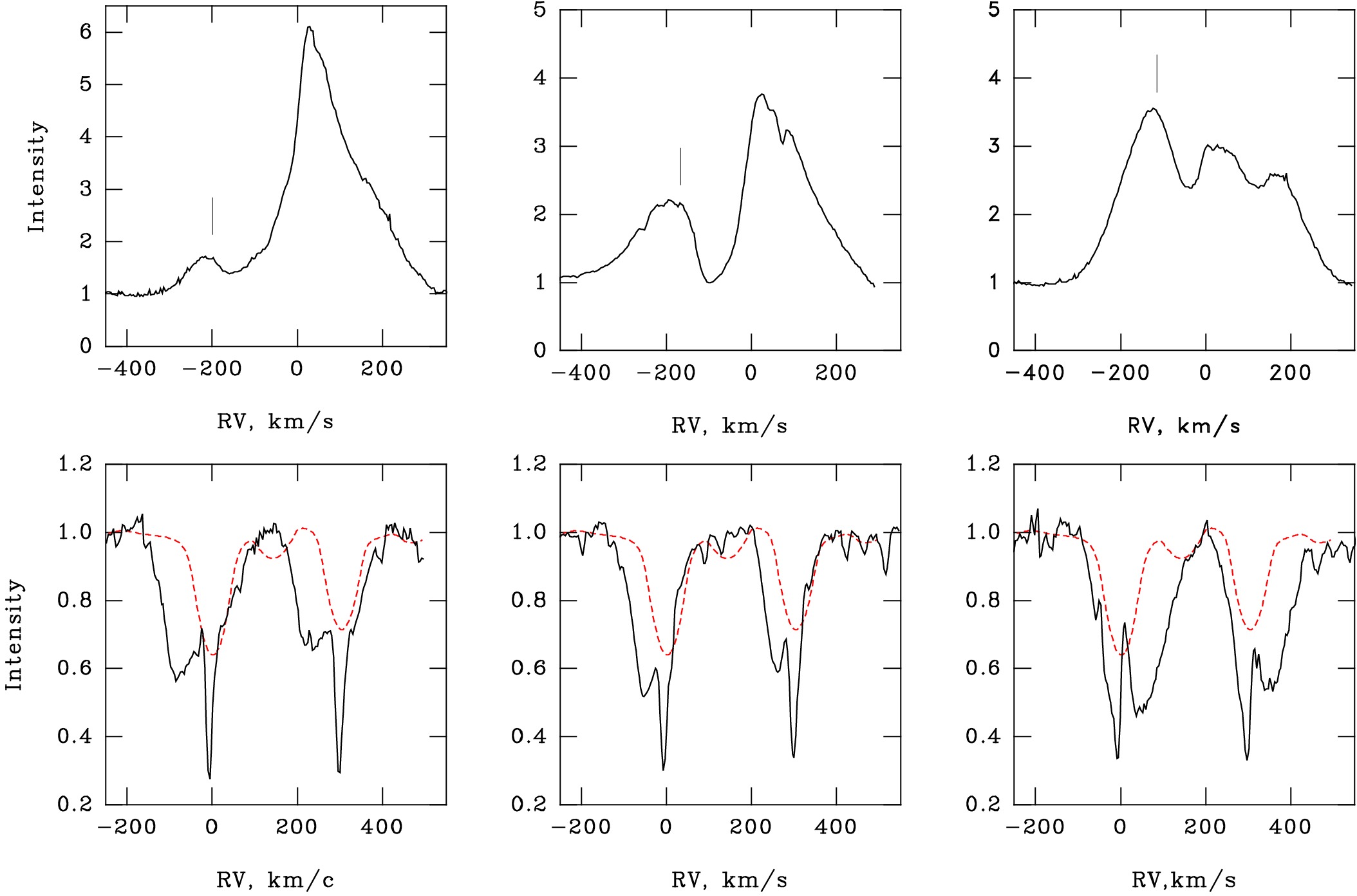}
\caption{\rm \footnotesize {Correlated variations in the H$\alpha$ (top) and Na D (bottom) profiles. The photospheric profile of the star 35 Leo (G1.5 IV-V)
rotationally broadened to $v\,\sin i$ = 50 km\,s$^{-1}$ (dashed line) is superimposed on the Na D profile.  }}
\end{figure}
%%%%%%%%%%%%%%%%%%%%%%%%%%%%%%%%%%%%%%%%%%%%%%%%%% 
          
%%%%%%%%%%%%%%%%%%%%%%%%%%%%%%%%%%%%%%%%%%%%%%%%%%%%%%%%%%%%%%%                
\begin{figure}[h!]
\epsfxsize=6.5cm
\vspace{0.6cm}
\hspace{5cm}\epsffile{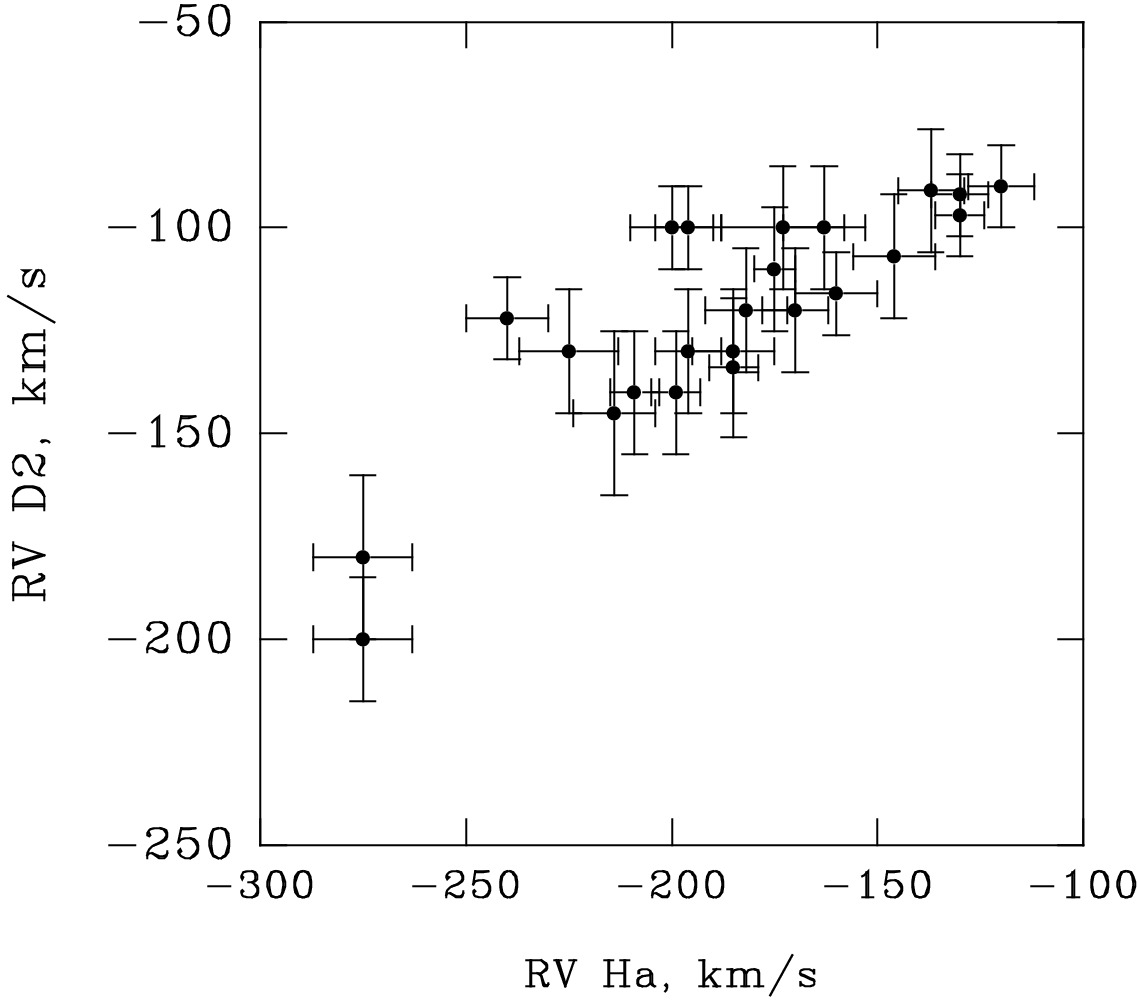}
\caption{\rm \footnotesize {Comparison of radial velocities of wind measured from the H$\alpha$ and Na D2 profiles. }}
\end{figure}
%%%%%%%%%%%%%%%%%%%%%%%%%%%%%%%%%%%%%%%%%%%%%%%%%%%%%%%%%%%%%%%      
                                                                 
The H$\alpha$ and Na D profiles are compared in Fig. 3.                
It can be seen that the additional (to the photospheric
profile) absorption in the Na D lines corresponds to
the absorption dips in the blue or red H$\alpha$ wing. The
radial velocity of the wind can be measured from the
blue edge of the absorption in the Na D lines, but the
noise at the continuum level does not always allow
this edge to be determined. In those cases where
this was possible, we measured the radial velocity of the
wind from the Na D2 line and from the blue edge of
the absorption in the H$\alpha$ line indicated by the vertical
dashes in Fig. 3. The result is shown in Fig. 4. A                 
clear correlation between the velocities measured by
the two methods confirms our assumption that the
dips in the H$\alpha$ wing are actually absorptions. The
radial velocity of the wind measured in all spectra from
the H$\alpha$ line is given in Table~1.
The three most characteristic types of H$\alpha$ profile
that alternate or pass into one another can be identified
(Fig. 5). They can be arbitrarily designated                        
as “wind and accretion” (II B + II R), “disk wind”
(II B), and “fast wind” (IV B or P Cyg). Out of the
32 observing nights, the “disk wind” (14 nights) and
“wind and accretion” (11 night) were observed most
frequently.

%%%%%%%%%%%%%%%%%%%%%%%%%%%%%%%%%%%%%%%%%%%%%%%%%%%%%%
\begin{figure}[h!]
\epsfxsize=15cm
\vspace{0.6cm}
\hspace{1cm}\epsffile{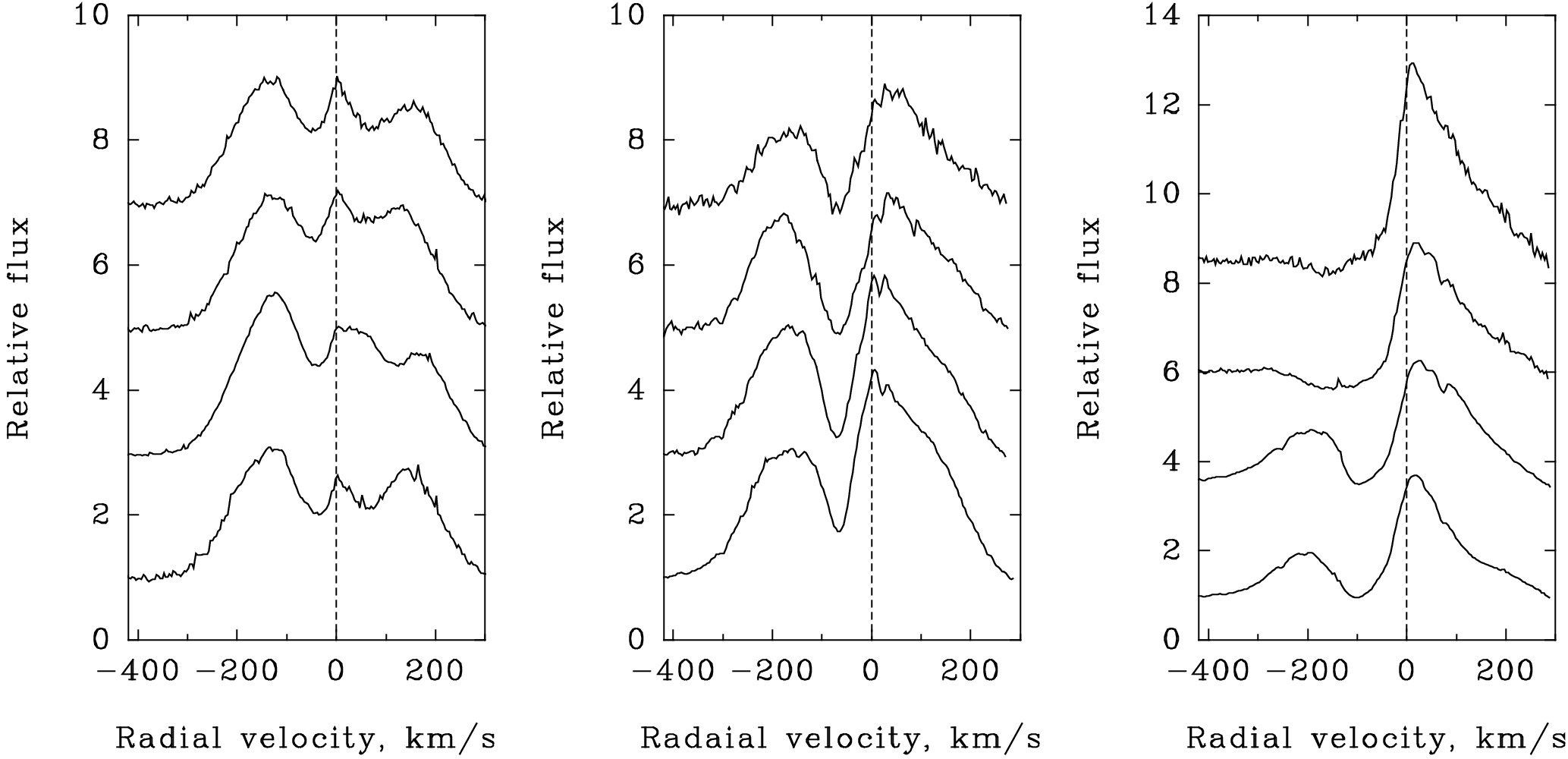}
\caption{\rm \footnotesize {Seies of spectra showing three types of H$\alpha$ profiles.  The left panel presents the profiles with signatures of wind and
accretion. The dates at which the spectra were taken, from top to bottom: March 7, 8, 9, and 10, 2015. The middle panel
presents the profiles typical of a disk wind (March 25, 26, 27, and 31, 2014). The right panel presents the series of spectra
including the “fast wind” episode (October 2, 3, 4, and 5, 2014).  }}
\end{figure}
%%%%%%%%%%%%%%%%%%%%%%%%%%%%%%%%%%%%%%%%%%%%%%%%%%%%%%%%%%

How fast do the line profiles change? Fig. 5
shows examples of several short series of spectra,
each taken on 4–6 consecutive nights. The profile
may not change for four days, i.e., there is no rotational
modulation, the type of profile does not depend
on which side of the star we see. Both the appearance
and disappearance of the “fast wind” (P Cyg profile)
occur rapidly, in one day. Out of the 32 observing
nights, P Cyg profiles were observed in seven cases.
This means that at an average duration of the continuous
series of about three consecutive nights, the
probability of detecting a P Cyg profile is about 0.2.
Consequently, it can be assumed that, on average,
such a profile appeared approximately once in 15 days,
i.e., once in several stellar rotations.

The main result of our observations is the detection
of a correlation between the stellar brightness
and the radial velocity of the wind (Fig. 6). The same        
correlation is shown in Fig. 7 as a gray-scale map. The       
entire brightness range was divided into 0$^m$.2 bins, and
the spectra corresponding to each brightness bin were
averaged.

%%%%%%%%%%%%%%%%%%%%%%%%%%%%%%%%%%%%%%%%%%%%%%%%%%%%%%%%%%%%%%%
\begin{figure}[h!]
\epsfxsize=8cm
\vspace{0.6cm}
\hspace{4cm}\epsffile{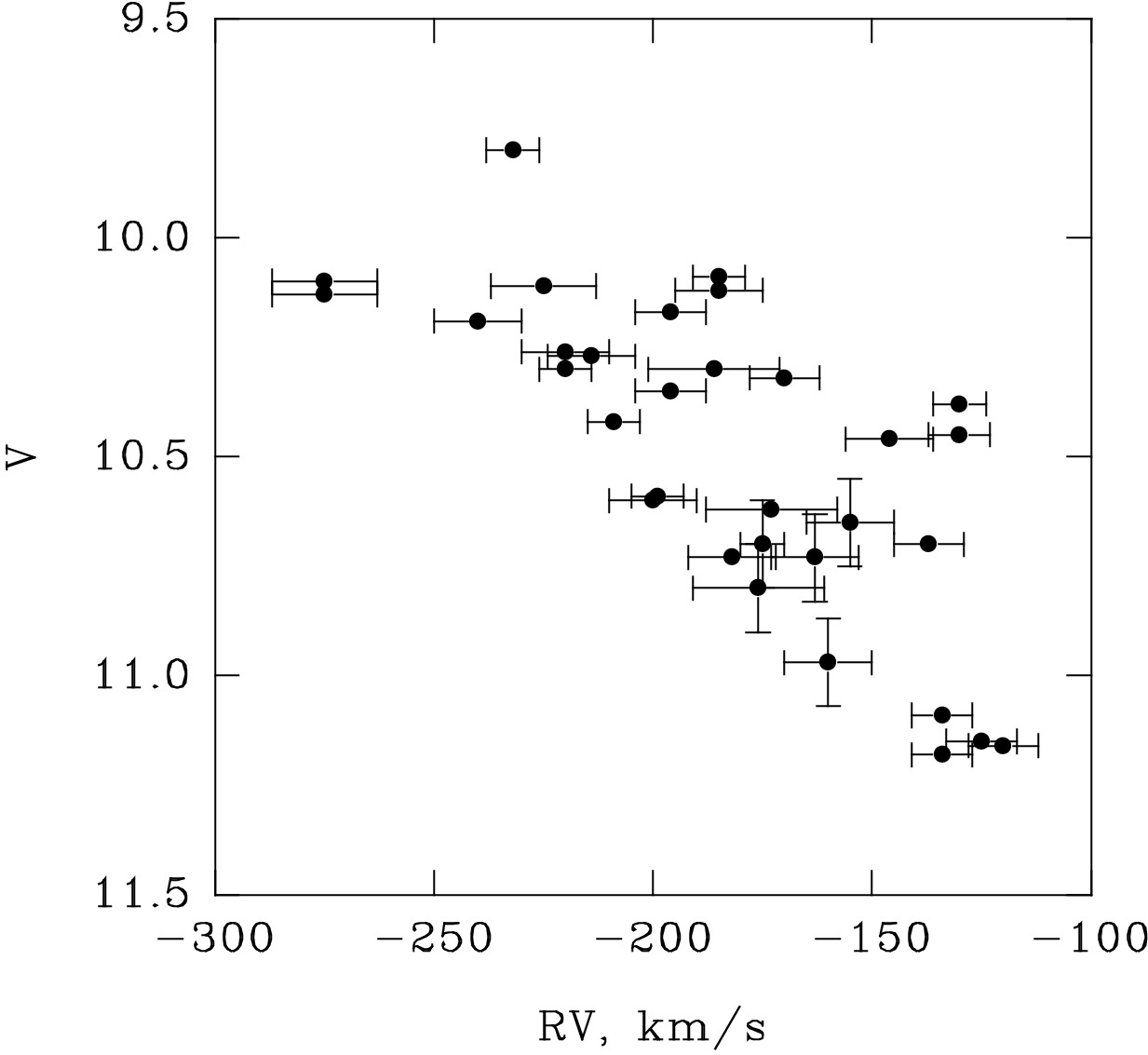}
\caption{\rm \footnotesize {Correlation between the stellar brightness $V$ and the radial velocity of the wind $RV$. }}
\end{figure}
%%%%%%%%%%%%%%%%%%%%%%%%%%%%%%%%%%%%%%%%%%%%%%%%%%%%%%%%%%
\begin{figure}[h!]
\epsfxsize=10cm
\vspace{0.6cm}
\hspace{2cm}\epsffile{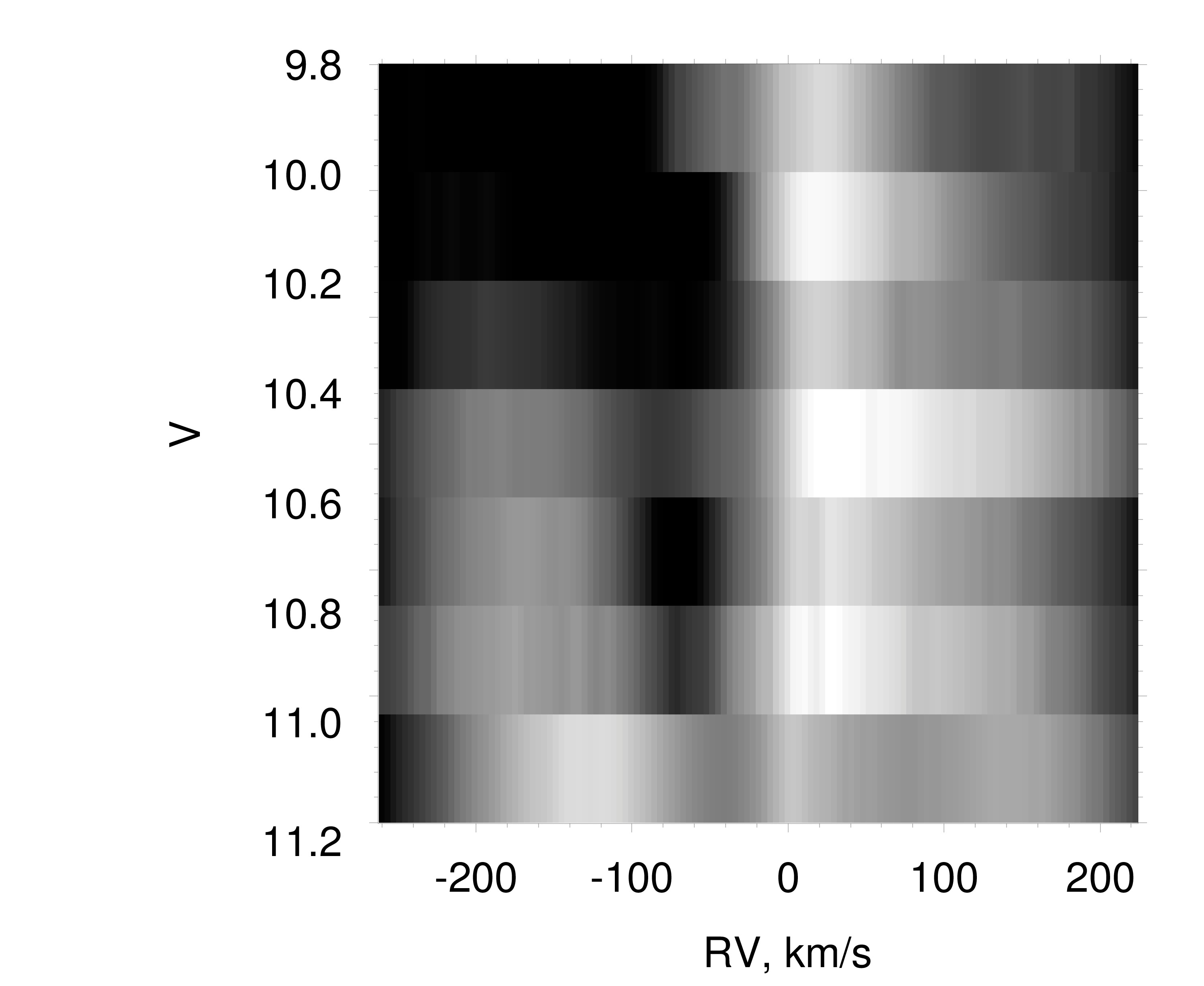}
\caption{\rm \footnotesize {Gray-scale map of H$\alpha$ profiles at various stellar brightness levels. The white and black areas correspond to the maximum and
minimum intensities in the H$\alpha$ profile.  }}
\end{figure}
%%%%%%%%%%%%%%%%%%%%%%%%%%%%%%%%%%%%%%%%%%%%%%%%%%%%%%%%%%%%

The rates of change of the stellar brightness $V$ and the radial 
velocity of the wind $RV$ can be characterized as follows. 
The brightness changes in one day by 0$^m$.1--0$^m$.2 in most cases 
and by as much as 0$^m$.45 in rare cases. The radial velocity of the wind
changes in one day by 10--40 km\,s$^{-1}$ in most cases
and by 90 km\,s$^{-1}$ in rare cases. The entire amplitude of
the brightness and velocity variations shown in Fig. 6
refers to long time intervals: about one month. As
has been pointed out above, previous studies have
shown that the brightness variations in RY Tau are
caused mainly by circumstellar extinction variations.
Our observations also confirm this conclusion. If
the $V$ brightness of the star changed due to accretion,
i.e., the appearance of hot spots on the stellar
surface, then the depth of photospheric lines in the
range 5000--6000\,\AA\, should have changed because
of the veiling effect. As the brightness changes by one
magnitude, the line depth must decrease accordingly
by a factor of 2.5. However, our observations show
that this effect is completely absent. At a difference
in stellar brightness by 1$^m$, the depth of photospheric
lines does not change within the measurement accuracy,
$\pm$2\% of the continuum level. Consequently, the
brightness variations in the visible spectral range are
caused mainly by the obscuration of the star by dust.

From the variations in the equivalent widths of
emission lines, we can conclude how vast the region
eclipsed by the dust screen is. In Fig. 8, the equivalent           
widths of the H$\alpha$ emission line and the [OI] 6300\,\AA\,
emission are plotted against the stellar brightness.
The H$\alpha$ emission is formed in the immediate vicinity
of the star, in the magnetosphere and the wind (Kurosawa
et al. 2011), as well as on the stellar surface
at the base of the accretion columns (Dodin 2015).
The [OI] 6300\,\AA\, emission line peak in the spectrum
of RY Tau is at a radial velocity of about -9 km\,s$^{-1}$
relative to the star. This line is formed in the rarefied
gas of a collimated wind (jet). The equivalent width
of the [OI] emission was measured in the range from
-30 to +30 km\,s$^{-1}$ relative to the peak. The [OI]
emission originating in the Earth's atmosphere is
usually shifted due to the velocity difference and does
not affect these measurements.
In the case where only the stellar photosphere is
screened, the emission line fluxes remain constant.
In this case, the emission line equivalent width will
increase with decreasing brightness as is indicated by
the dashed line in Fig. 8. Our observations show that
the equivalent width of both emission lines actually
increases with decreasing stellar brightness. This
means that the emission region is obscured incompletely
by dust. In addition, a significant intrinsic
variability in the lines is observed. In particular,
the range of variations in the [OI] 6300\,\AA\, line is
noticeably greater than might be expected when the
brightness changes by one magnitude. The formation
region of this line closest to the star probably also
responds to changes in the magnetosphere: the line
intensity decreases when a ''fast wind'' appears.

%%%%%%%%%%%%%%%%%%%%%%%%%%%%%%%%%%%%%%%%%%%%%%%%%%%%%%%
\begin{figure}[h!]
\epsfxsize=12cm
\vspace{0.6cm}
\hspace{2cm}\epsffile{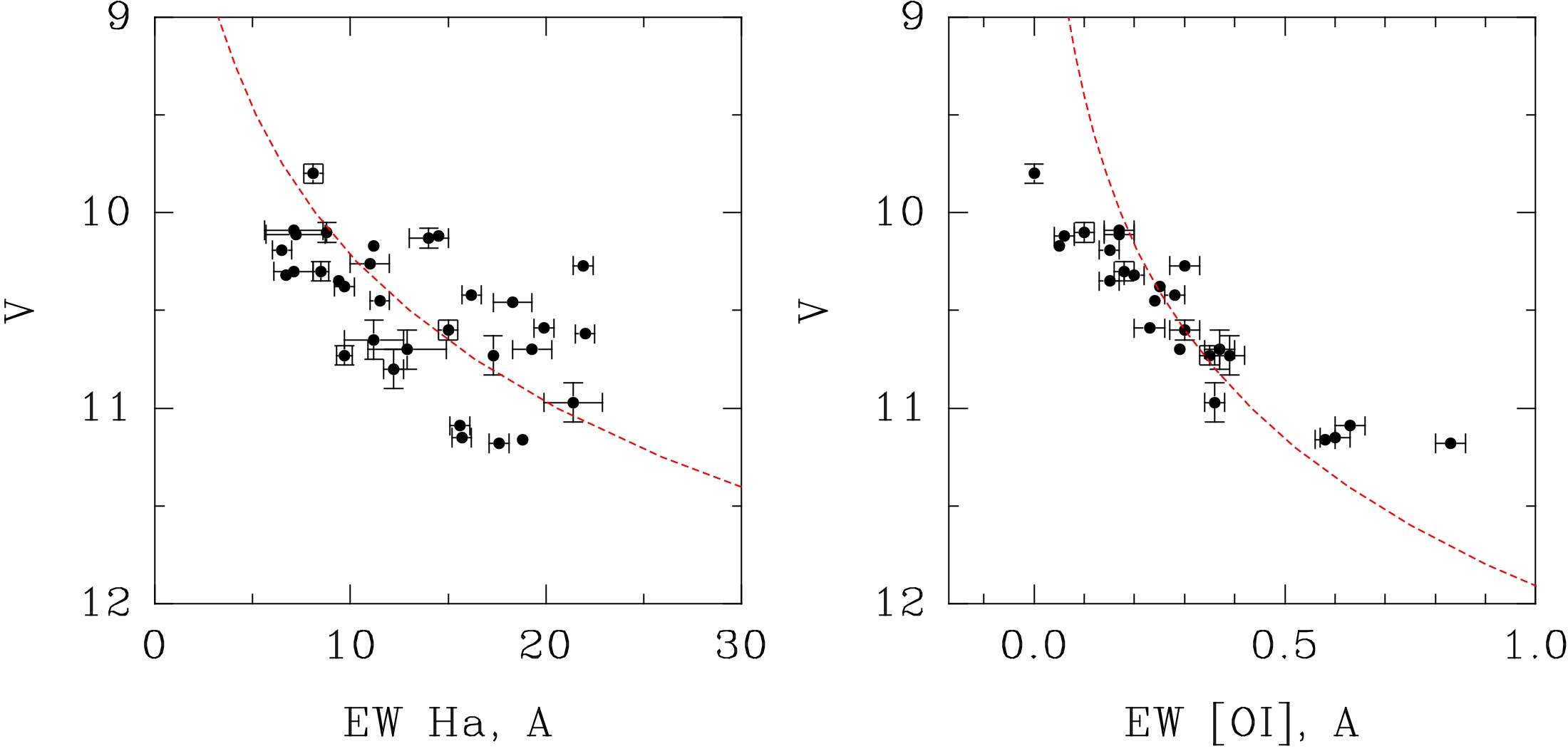}
\caption{\rm \footnotesize {Equivalent widths (EW) of the H$\alpha$ and [OI] 6300\,\AA\, emission lines versus stellar brightness (V). The dashed line
indicates the expected dependence in the case of a constant line flux.  }}
\end{figure}
%%%%%%%%%%%%%%%%%%%%%%%%%%%%%%%%%%%%%%%%%%%%%%%%%%%%%%%%%%%%%%%%%

\subsection*{Discussion}

In this paper, we investigate the dynamics of the
stellar wind from RY Tau. The blueshifted absorption in the H$\alpha$ 
and Na D line profiles is supposed to be an indicator of the wind flows.  
It should be mentioned, that an alternative interpretation of the H$\alpha$ 
line profiles was proposed by Grinin and Tambovtseva (1995) when studying 
the variability of emission line profiles in Herbig Ae/Be stars: if a dust cloud obscuring successively different parts of the emission region from the
observer moves around the star, then characteristic
changes must be observed in the H$\alpha$ line profile.
However, the views of the causes of spectroscopic and
photometric variability of T Tauri stars have changed
noticeably over the elapsed 20 years. Great progress
has been achieved in studying the interaction of the
stellar magnetic field with the accretion disk and the
formation of a wind and accretion flows. In particular,
these processes were shown to be essentially
nonstationary on a time scale of several stellar rotations
(for a review, see Romanova et al. 2014). This
inevitably affects the observed emission line profiles. 
There is good reason to believe that the
variability of emission lines in the spectra of classical
T Tauri stars reflects the actual dynamical processes
and is not caused only by eclipses. For example, fast
irregular variability of the Balmer lines, including the
episodic appearances of P Cyg profiles indicating active mass ejections, 
has been repeatedly observed
in the T Tauri star SU Aur, which in many ways
resembles RY Tau (Jones and Basri 1995).

Cyclic rearrangements of the magnetosphere accompanied
by coronal mass ejections can in principle
explain the observed variability in the H$\alpha$ profile,
where the ''quiescent wind'' is occasionally disrupted
by ''fast wind'' episodes. A similar effect is observed in
the star AA Tau, which is aslo oriented nearly equator-on
(Bouvier et al. 2003). The synchronous changes in the radial 
velocity of the wind and accretion flows 
were interpreted in terms of the model
of a dynamical rearrangement of the magnetosphere.
During the magnetospheric inflation, the slope of the
magnetic field lines in the inner disk changes and, accordingly,
the radial velocity of the gas moving along
the field lines changes. For AA Tau, the observed
brightness variations are periodic. It was interpreted as
recurrent occultation of the star by the warped inner disk 
due to the inclined magnetosphere (Bouvier et al. 1999)

In contrast to AA Tau, RY Tau shows no rotational
modulation either in the brightness variations or in the
emission line intensity or profile variations. Since
there is no veiling of the photospheric spectrum in
the visible spectral range, one can hardly expect any
photometric variability caused by nonstationary accretion.
The magnetosphere of RY Tau is probably
inclined only slightly to the rotation axis and is axially
more symmetric than that in other T Tauri stars;
therefore, we have a rare opportunity to investigate
the dynamical phenomena in the magnetosphere and
the wind undistorted by rotational modulation.
In RY Tau, we observe a new effect, the wind
interaction with the dusty environment of the star.
The correlation between the wind velocity and circumstellar
extinction suggests that the distance from
the ''fast wind'' formation region to the dust screen
region is small, 0.1--0.2 AU, i.e., the dust screen is
at the inner rim of the dust disk. In other words, the
irregular photometric variability of RY Tau is governed
not by random eclipses by dust clouds in the accretion
disk atmosphere but by MHD processes in the stellar
magnetosphere. Presumably, the magnetospheric
inflation and the succeeding gas ejection temporarily
destroy the dust screen at the inner disk boundary.
After the restoration of the magnetosphere, the disk
wind again lifts dust and restores the dust screen.

Amplitude of light varibility, caused by extinction due to dust in the disk wind of RY Tau,
is $\sim$ 1$^m$. The mass of the dust causing such extinction can be estimated.
To attenuate the light from the star by one magnitude, the number of dust particles on the line of sight in a column with a cross section of 1 cm$^2$ must be, in order of magnitude
	$\approx 1/ \left(\pi\cdot a^2\right)$, where a - radius of a particle.
Then, the dust column density is		
		$m_{col} \approx 4/3\cdot a \cdot \rho$, where $\rho$ is average density (g/cm$^3$).	
In projection onto a star of radius $R_*$ the mass of the dust particles is
		$m_{col}\cdot \pi \cdot R_*^2$. 		
At an inner radius of the dusty disk $R_{in}$  the mass of dust in a ring zone around the star
of height   $2 \cdot R_*$ and the length of $2\cdot\pi\cdot R_{in}$, is
    $4 \cdot \left(R_{in}/R_*\right) \cdot m_{col} \cdot \pi \cdot R_*^2$. 		

At a tangential (across the line of sight) velocity  $V_t$, the dusty wind travels a distance equal to the diameter of the star, in the time   $t = 2 R_*/ V_t$.
Consequently, the dust mass flux is
\[\dot{M}_{dust} \approx 4 \cdot \left(R_{in} / R_*\right) \cdot m_{col} \cdot\pi\cdot R_*^2 / t,\]
or  
\[\dot{M}_{dust} \approx 8 \cdot R_{in} \cdot a \cdot \rho \cdot V_t.\]

Assuming, as an example,  
$a = 0.3$ $\mu$m,  $\rho = 2$ g/cm$^3$,  $R_{in} = 0.2$ AU  and  $V_t = 100$ km/s, we obtain  
 $\dot{M}_{dust} \approx 2 \cdot 10^{-11}$ $M_\odot$/yr.
At the dust to gas mass ratio 0.01,  the mass loss rate in the wind is 
$\dot{M}_{wind} \approx 2\cdot 10^{-9}$ $M_\odot$/yr.

Some uncertainty in this estimate may stem from
the fact that we do not know the direction of the disk
wind velocity vector. Only the radial velocity of the
wind is determined from observations. For comparison,
the most reliable value of the {\it accretion} rate onto
RY Tau derived from the veiling level in the ultraviolet
spectral range is  
$\dot{M}_{accr}$ = (6--9)$\cdot10^{-8}$ M$_\odot$ yr$^{-1}$
(Calvet et al. 2004), i.e., an order of magnitude higher
than the outflow rate $\dot{M}_{wind}$.

The variability of young stars due to circumstellar
extinction shows itself most clearly in UX Ori
stars, which are more massive and hotter than T Tauri
stars. Since the inner boundary of the dust disk is
farther from the star (about 0.5 AU), one might expect
the correlation between the wind and circumstellar
extinction in such stars to be not so obvious. In
T Tauri stars, the boundary of the dust disk is closer
to the region from where the ''magnetospheric'' wind
starts; therefore, these effects are more pronounced.

The variability mechanism considered here concerns
the short-term brightness variations on a time
scale of several stellar rotations. The longer-term
dimming events can be caused by a nonuniform distribution
of dust in the accretion disk.
In course of disk accretion, local concentrations of dust arrive gradually at the inner boundary of the disk, from where the dust gets into the wind and causes a deep and long-lasting event of extinction.
Such an effect was observed in the T Tauri star RW Aur A. During a deep dimming of the star
in the optical range, its near-infrared (2–5 $\mu$m) brightness rose. This was interpreted as a manifestation of hot dust in the wind directed toward the observer (Shenavrin et al. 2015).

\subsection*{Conclusion}
The photometric variability of RY Tau is attributable
to circumstellar extinction variations. The
dust screen responsible for the stellar variability lies
at the inner boundary of the dust disk at a distance of
about 0.2 AU from the star. The radial velocity of the
wind is, on average, $\sim$\,100 km\,s$^{-1}$, but gas ejections
a higher velocity, up to 280 km\,s$^{-1}$, are occasionally
observed. A clear correlation between the radial
velocity of the wind and the stellar brightness has
been found for the first time. We assume that
the circumstellar extinction variations result from a
cyclic rearrangement of the stellar magnetosphere
and coronal mass ejections, which affects the dusty
disk wind near the inner boundary of the disk.

\subsection*{Acknowledgements}
We thank K.N. Grankin for providing the photometric data for Fig. 1.

\newpage

%%%%%%%%%%%%%%%%%%%%%%
\subsection*{References}
             
\noindent 
Agra-Amboage V., Dougados C., Cabrit S., et al. 2009,  A\&A, 493, 1029 
%% [O I] sub-arcsecond study of a microjet from an intermediate mass young star: RY~Tauri

\noindent
Akeson R.L.,  Walker C.H.,  Wood K.,  et al.  2005, ApJ, 622, 440.
%% Observations and Modeling of the Inner Disk Region of T Tauri Stars

\noindent
Blandford R.D. \& Payne D.G. 1982, MNRAS, 199 883
%% Hydromagnetic flows from accretion discs and the production of radio jets

\noindent
Bouvier J., Chelli A., Allain S., et al. 1999, A\&A, 349, 619 
%% warped disk in AA Tau

\noindent
Bouvier J., Grankin K.N., Alencar S.H.P., et al. 2003, A\&A, 409, 169 
%% Eclipses by circumstellar material in the T Tauri star AA Tau. II. Evidence for non-stationary magnetospheric accretion

\noindent
Bouvier J.,  Alencar S.H.P., Harries  T.J., et al. 2007, Protostars \& Planets V (Ed. 

 B. Reipurth,  D. Jewitt, and K. Keil,  University of Arizona Press, Tucson, 2007, p. 479
%% Magnetospheric Accretion in Classical T Tauri Stars (warp disk in AA Tau)

\noindent 
Calvet N., Muzerolle J., Briceno C., et al. 2004, AJ, 128, 1294 
%% The Mass Accretion Rates of Intermediate-Mass T Tauri Stars

\noindent 
Camenzind M. 1990, Reviews in Modern Astronomy , 3, 234 
%% Magnetized Disk-Winds and the Origin of Bipolar Outflows

\noindent 
Chou M.-Y.,  Takami M.,  Manset N.,  et al. 2013, ApJ , 145, 108 
%% Time Variability of Emission Lines for Four Active T Tauri Stars. I. October-December in 2010

\noindent 
Coffey D., Dougados C., Cabrit S., et al. 2015, ApJ , 804, 2 
%% A Search for Consistent Jet and Disk Rotation Signatures in RY~Tau

\noindent
Dodin A.V. 2015, Astron. Lett., 41, 196 
%%Non-LTE modeling of the structure and spectra of hot accretion spots on the surface of young stars

\noindent 
Ferreira J.,  Pelletier G.,  Appl S. 2000,  MNRAS , 312, 387 
%% Reconnection X-winds: spin-down of low-mass protostars

\noindent
Goodson A.P., Winglee R.M. , Bohm K.-H. 1997,  ApJ, 489, 199 
%% Time-dependent Accretion by Magnetic Young Stellar Objects as a Launching Mechanism for Stellar Jets

\noindent
Grinin V.P., Kiselev N.N., Chernova G.P., et al. 1991, Astrophys. \& Space Sci., 186, 283
%%The investigations of 'zodiacal light' of isolated AE-Herbig stars with nonperiodic algol-type minima

\noindent
Grinin V.P. \& Tambovtseva L.V. 1995, A\&A,  293, 96
%Variable circumstellar obscuration and variability of emission lines in the spectra of the Herbig Ae/Be stars.

\noindent
Grinin V.P. \& Tambovtseva L.V. 2011, Astronomy Reports, 55, 704 
%% Дисковый ветер в излучении молодых звезд промежуточных масс

\noindent
Gullbring E.,  Hartmann L. , Briceno C.,  et al. 1998, ApJ, 492, 323 
%% Disk Accretion Rates for T Tauri Stars

\noindent 
Hartigan P.,  Edwards S.,  Ghandour L. 1995,  ApJ , 452, 736 
%% Disk Accretion and Mass Loss from Young Stars

\noindent 
Herbst W. \&  Stine P.C. 1984, AJ, 89, 1716 
%% Photometric variations of Orion population stars. III - RY~Tau, T Ori, NV Ori, and HH AUR

\noindent 
Herbst W.,  Herbst D.K.,  Grossman E.J.,  et al. 1994,  AJ, 108, 1906 
%% Catalogue of UBVRI photometry of T Tauri stars and analysis of the causes of their variability

\noindent 
Isella A., Carpenter J.M., Sargent A.I. 2010,  ApJ , 714, 1746 
%% Investigating Planet Formation in Circumstellar Disks: CARMA Observations of RY~Tau and Dg Tau

\noindent
Jones C.M. \& Basri G. 1995, ApJ, 449, 341
%% The line profile variability in SU Aur

\noindent 
Koenigl A. 1991, ApJ , 370, L39 
%% Disk accretion onto magnetic T Tauri stars

\noindent 
Kurosawa R., Romanova M.M., Harries T.J. 2011, MNRAS , 416, 2623 
%% Multidimensional models of hydrogen and helium emission line profiles for classical T Tauri stars: method, tests and examples

\noindent 
Kurosawa R. \& Romanova M.M. 2012, MNRAS , 42, 2901 
%Line formation in the inner winds of classical T Tauri stars: testing the conical-shell wind solution

\noindent 
Lopez-Martinez F. \& Gomez de Castro, A.I. 2014, MNRAS , 442, 2951 
%% Constraints to the magnetospheric properties of T Tauri stars - I. The C II], Fe II] and Si II] ultraviolet features

\noindent 
Matt S. \& Pudrits R.E. 2005,  ApJ , 632, L135 
%% Accretion-powered Stellar Winds as a Solution to the Stellar Angular Momentum Problem

\noindent 
Petrov P.P., Zajtseva G.V., Efimov Yu.S., et al. 1999, A\&A , 341, 553 
%% Brightening of the T Tauri star RY~Tauri in 1996. Photometry, polarimetry and high-resolution spectroscopy

\noindent 
Petrov P.P., Gahm G.F., Gameiro J.F., et al. 2001, A\&A , 369, 993 
%% Non-axisymmetric accretion on the classical TTS RW Aur A

\noindent 
Petrov P.P., Gahm G.F., Stempels H.C., et al. 2011, A\&A , 535, 6 
%% Accretion-powered chromospheres in classical T Tauri stars

\noindent 
Pott J.-U.,  Perrin M.D.,  Furlan E., et al. 2010, ApJ , 710, 265 
%% Ruling Out Stellar Companions and Resolving the Innermost Regions of Transitional Disks with the Keck Interferometer

\noindent 
Reipurth B., Pedrosa A.,  Lago M.T.V.T. 1996,  A\&A Suppl. Ser., 120, 229 
%% Ha emission in pre-main sequence stars. I. an atlas of line profiles.

\noindent 
Robitaille T.P., Whitney B.A., Indebetouw R., et al. 2007, ApJSS., 169, 328 
%% Interpreting Spectral Energy Distributions from Young Stellar Objects. II. Fitting Observed SEDs Using a Large Grid of Precomputed Models

\noindent 
Romanova M.M., Ustyugova G.V., Koldoba A.V., et al. 2009, MNRAS, 399, 1802 
%% Launching of conical winds and axial jets from the disc-magnetosphere boundary: axisymmetric and 3D simulations

\noindent 
Romanova M. M.,  Lovelace R. V. E.,  Bachetti M., et al. 2014, Physics at the magnetospheric %% перенос строки!

  boundary (Ed. E. Bozzo et al., Geneva, Switzerland, EPJ Web of Conf., 2014), p. 64.
%% MHD simulations of magnetospheric accretion, ejection and plasma-field interaction

\noindent 
Safier P.N.,  ApJ 1993, 408, 115 .
%% Centrifugally driven winds from protostellar disks. I - Wind model and thermal structure

\noindent 
Schegerer A. A.,  Wolf S.,  Ratzka Th.,  et al. 2008, A\&A, 478, 779 
%% The T Tauri star RY~Tauri as a case study of the inner regions of circumstellar dust disks

\noindent 
Shenavrin V.I.,  Petrov P.P.,  Grankin K.N. 2015, IBVS, 4628 
%% Hot Dust Revealed During the Dimming of the T Tauri Star RW Aur A

\noindent 
Shu F.,  Najita J.,  Ostriker E.,  et al. 1994, ApJ, 429, 781 
%% Magnetocentrifugally driven flows from young stars and disks. 1: A generalized model

\noindent 
Siess L.,  Dufour E.,  Forestini M. 2000, A\&A, 358, 593 
%% An internet server for pre-main sequence tracks of low- and intermediate-mass stars

\noindent 
Siwak M.,  Rucinski S.M., Matthews J.M., et al. 2011, MNRAS, 415, 1119

\noindent 
Skinner S.L.,  Audard M.,  Gudel M. 2011,  ApJ, 737, 19 
%% Chandra Evidence for Extended X-Ray Structure in RY~Tau

\noindent 
St-Onge G.,  Bastien P.,  ApJ 2008, 674, 1032 
%% A Jet Associated with the Classical T Tauri Star RY~Tauri

\noindent
Tambovtseva L.V. \&  Grinin V.P. 2008,  Astron. Lett. 34, 231 
%% Пыль в дисковых ветрах малодых звезд как источник околозвездной экстинкции

\noindent 
White R.J. \& Ghez V. 2001,  ApJ, 556, 265 
%% Observational Constraints on the Formation and Evolution of Binary Stars

\noindent
Zaitseva G.V. 2010, Astrophysics, 53, 212 
%%Анализ 30-летнего ряда фотоэлектрических наблюдений RY Тельца. I. Поиск возможных периодичностей

\noindent
Zajtseva G., Petrov P. Ilyin  I., et al. 1996, IBVS, 4408
%% RY~Tauri at High Brightness
\\
\\
{\it Accepted for publication in Astronomy Letters, 2016, Vol. 42, No. 3, pp. 193–203.}
\end{document}